\documentclass[prd,nofootinbib,showpacs,preprint]{revtex4}
\usepackage{amsmath}
\usepackage{graphicx}
\graphicspath{{Figs/}}
\usepackage{dcolumn}
\usepackage{bm}
\usepackage{amssymb}
\usepackage[usenames,dvipsnames]{color}
\usepackage{slashed}
\usepackage[dvipdfm,colorlinks,citecolor=blue]{hyperref}

\begin{document}
\title{Effects of Planck Scale Physics on Neutrino Mixing Parameters in Left-Right Symmetric Models}
\author{Debasish Borah}
\email{dborah@tezu.ernet.in}
\affiliation{Department of Physics, Tezpur University, Tezpur-784028, India}

\begin{abstract}
Left right symmetric models (LRSM) are extensions of the standard model by an enlarged gauge group $SU(2)_L \times SU(2)_R \times U(1)_{B-L}$ where automatic inclusion of right handed fermions as $SU(2)_R$ doublets guarantees a natural seesaw origin of neutrino masses. Apart from the extended gauge symmetry, LRSM also has an in-built global discrete symmetry, called D-parity which ensures equal gauge couplings for left and right sectors. Motivated by the fact that global symmetries are expected to be explicitly broken by theories of quantum gravity, here we study the effects of such gravity or Planck scale physics on neutrino masses and mixings by introducing explict D-parity breaking Planck scale suppressed higher dimensional operators. Although such Planck scale suppressd operators have dimension at least six in generic LRSM, dimension five operators can also arise in the presence of additional scalar fields which can be naturally accommodated within $SO(10)$ grand unified theory (GUT) multiplets. We show that, such corrections can give rise to significant changes in the predictions for neutrino mixing parameters from the ones predicted by tree level seesaw formula if the left right symmetry breaking scale is lower than $10^{14}$ GeV.
\end{abstract}

\pacs{12.60.-i,12.60.Cn,14.60.Pq}
\maketitle
\section{Introduction}
Left-Right Symmetric Models (LRSM) \cite{lrsm} have been one of the most well-motivated extension of Standard Model (SM) studied in great details for last few decades. Apart from explaining the origin of parity breaking in weak interactions spontaneously, LRSM can also explain the origin of tiny neutrino masses \cite{PDG} naturally via seesaw mechanism \cite{ti, tii} without reference to very high scale physics such as grand unified theories (GUT). Supersymmetric versions of such models have several other motivations like protecting the scalar sector from quadratic divergences, providing a natural dark matter candidate among others. However, as studied previously in \cite{Mishra:2009mk, borah}, generic Supersymmetric Left-Right models are tightly constrained from consistent cosmology as well as successful gauge coupling unification point of view and in quite a few cases these models do not give rise to successful unification and consistent cosmology simultaneously. Recently, non-supersymmetric versions of LRSM were studied in the context of of gauge coupling unification and consistent cosmology \cite{borah12}. It was shown that minimal versions of LRSM can not give rise to unification and consistent cosmology simultaneously, but suitable extensions of these models can give rise to both of these desired properties and at the same time allowing the possibility of a low scale gauge symmetry.

Spontaneous breaking of exact discrete symmetries like parity (which we shall denote as
D-parity hereafter) has got cosmological
implications since they lead to frustrated phase transitions leaving behind a network of domain walls (DW). These 
domain walls, if not removed will be in conflict with the observed Universe \cite{Kibble:1980mv,Hindmarsh:1994re}. It was pointed 
out \cite{Rai:1992xw,Lew:1993yt} that Planck scale suppressed non-renormalizable operators 
can be a source of domain wall instability. The main theme of these works were to assume exact parity symmetry at tree level and introduce explicit parity breaking terms of higher order. As pointed out in \cite{Rai:1992xw}, any generic theories of quantum gravity should not respect global symmetries: both discrete and continous. Without worrying about the details of such symmetry breaking mechanism, our purpose is to study the effects of such terms which arise only in the form of higher dimensional operators. The role of such higher dimensional operators in destabilizing domain walls was studied in \cite{Mishra:2009mk, borah12}. Here we intend to study the effects of such operators on the neutrino sector, namely the neutrino mixing parameters. We find that in generic LRSM, such operators which affect neutrino paramaters can have dimension at least six and do not affect the neutrino masses and mixings significantly. However, in the presence of additional scalar fields, dimension five operators can arise and can affect the neutrino mass and mixings significantly. In particular, we incorporate the presence of gauge singlet scalar field which can naturally fit inside several $SO(10)$ GUT representations. As discussed in \cite{borah12}, such singlet extension of minimal LRSM also leads to domain wall disappearance which is not possible in the minimal versions. Here we study the effect of such higher dimensional operators on neutrino mixing parameters and find that the corrections can be very significant if the left right symmetry breaking scale is below $10^{14}$ GeV.

This paper is organized as follows. In section \ref{minLR} we discuss minimal LRSM with Higgs triplets and discuss how tiny neutrino mass arise in this model. In section \ref{HDLRSM} we discuss the possible higher dimensional and explicit parity breaking operators which can affect neutrino masses. Then in section \ref{num}, we present our numerical analysis on the effects of higher dimensional operators on neutrino mixing parameters and finally conclude in section \ref{conclude}.
\begin{figure}[ht]
\centering
\includegraphics[width=1.0\textwidth]{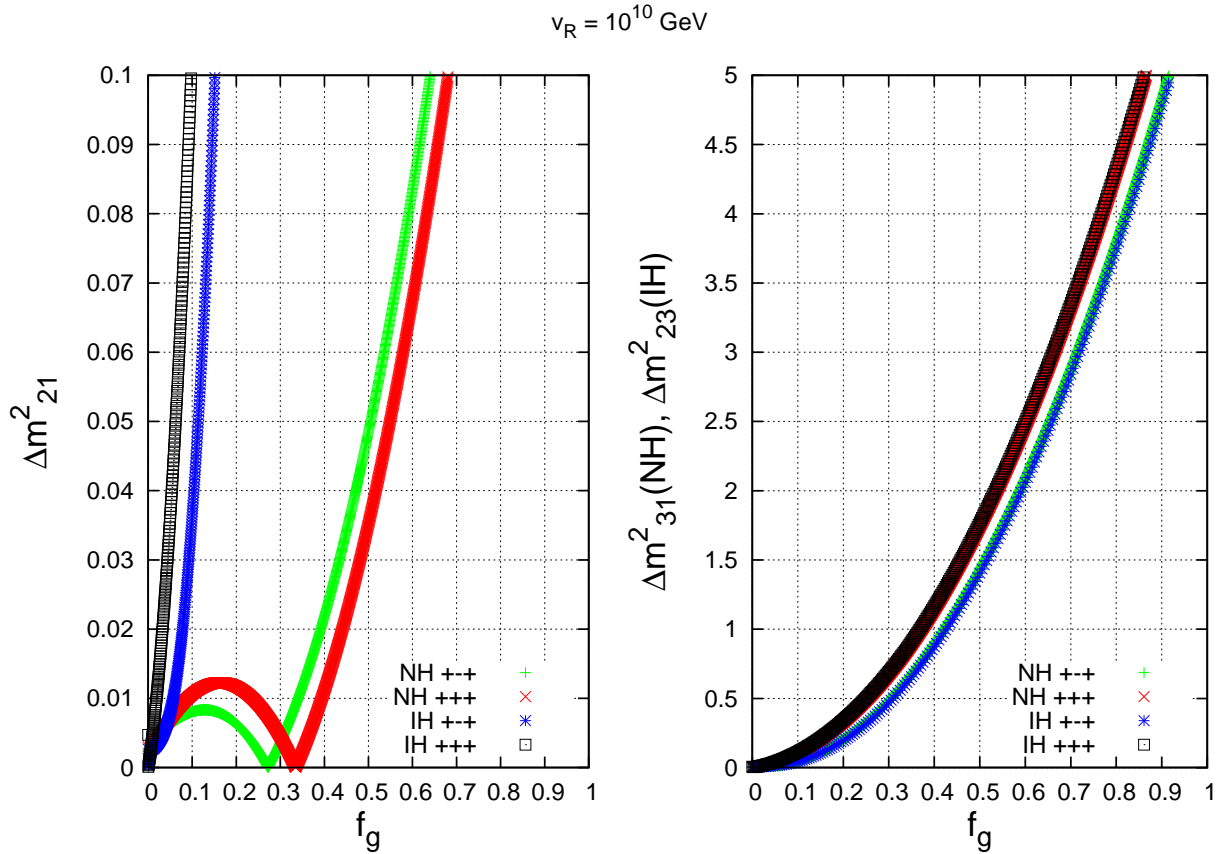}
\caption{Variations of $\Delta m^2_{21}$ and $\Delta m^2_{31}$(NH), $\Delta m^2_{23}$(IH) as a function of $f_g$ for $v_R = 10^{10}$ GeV}
\label{fig1}
\end{figure}
\begin{figure}[h]
\begin{center}
\includegraphics[width=1.0\textwidth]{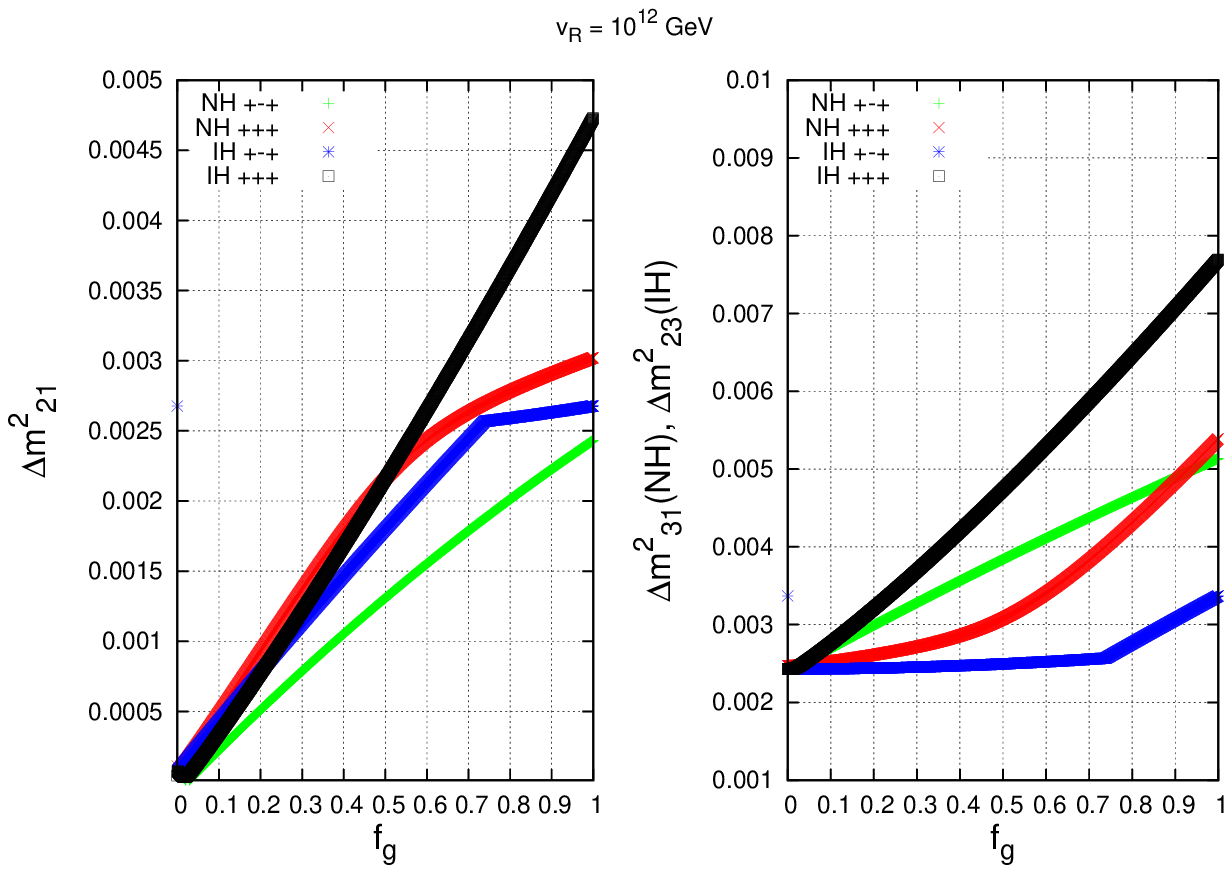}
\end{center}
\caption{Variations of $\Delta m^2_{21}$ and $\Delta m^2_{31}$(NH), $\Delta m^2_{23}$(IH) as a function of $f_g$ for $v_R = 10^{12}$ GeV}
\label{fig2}
\end{figure}
\begin{figure}[h]
\begin{center}
\includegraphics[width=1.0\textwidth]{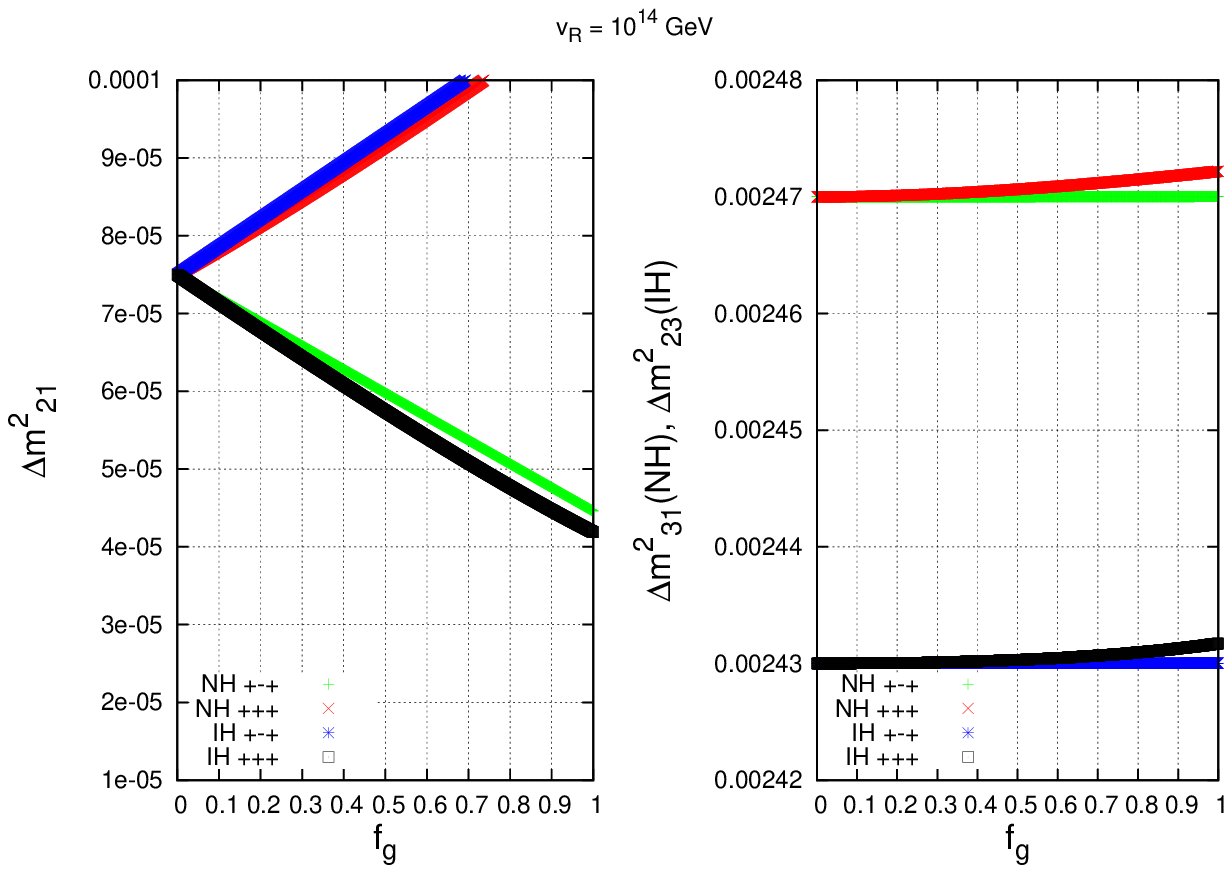} 
\end{center}
\caption{Variations of $\Delta m^2_{21}$ and $\Delta m^2_{31}$(NH), $\Delta m^2_{23}$(IH) as a function of $f_g$ for $v_R = 10^{14}$ GeV}
\label{fig3}
\end{figure}

\section{Neutrino Mass in LRSM}
\label{minLR}
The fermion content of minimal LRSM is
\begin{equation}
Q_L=
\left(\begin{array}{c}
\ u_L \\
\ d_L
\end{array}\right)
\sim (3,2,1,\frac{1}{3}),\hspace*{0.8cm}
Q_R=
\left(\begin{array}{c}
\ u_R \\
\ d_R
\end{array}\right)
\sim (3^*,1,2,\frac{1}{3}),\nonumber 
\end{equation}
\begin{equation}
\ell_L =
\left(\begin{array}{c}
\ \nu_L \\
\ e_L
\end{array}\right)
\sim (1,2,1,-1), \quad
\ell_R=
\left(\begin{array}{c}
\ \nu_R \\
\ e_R
\end{array}\right)
\sim (1,1,2,-1) \nonumber
\end{equation}
Similarly, the Higgs content of the minimal LRSM is
\begin{equation}
\Phi=
\left(\begin{array}{cc}
\ \phi^0_{11} & \phi^+_{11} \\
\ \phi^-_{12} & \phi^0_{12}
\end{array}\right)
\sim (1,2,2,0)
\nonumber 
\end{equation}
\begin{equation}
\Delta_L =
\left(\begin{array}{cc}
\ \delta^+_L/\surd 2 & \delta^{++}_L \\
\ \delta^0_L & -\delta^+_L/\surd 2
\end{array}\right)
\sim (1,3,1,2), \hspace*{0.2cm}
\Delta_R =
\left(\begin{array}{cc}
\ \delta^+_R/\surd 2 & \delta^{++}_R \\
\ \delta^0_R & -\delta^+_R/\surd 2
\end{array}\right)
\sim (1,1,3,2) \nonumber
\end{equation}
where the numbers in brackets correspond to the quantum numbers with respect to the gauge group $SU(3)_c\times SU(2)_L\times SU(2)_R \times U(1)_{B-L}$. In the symmetry breaking
pattern, the scalar $\Delta_R$ acquires a vacuum expectation value (vev) to break the gauge symmetry of LRSM into that of the standard model and then to $U(1)$ of electromagnetism by the vev of Higgs bidoublet $\Phi$:

$$ SU(2)_L \times SU(2)_R \times U(1)_{B-L} \quad \underrightarrow{\langle
\Delta_R \rangle} \quad SU(2)_L\times U(1)_Y \quad \underrightarrow{\langle \Phi \rangle} \quad U(1)_{em}$$

The relevant Yukawa couplings which leads to small non-zero neutrino
mass is given by
\begin{eqnarray}
{\cal L}^{II}_\nu &=& y_{ij} \ell_{iL} \Phi \ell_{jR}+ y^\prime_{ij} \ell_{iL}
\tilde{\Phi} \ell_{jR} +h.c.
\nonumber \\
&+& f_{ij}\ \left(\ell_{iR}^T \ C \ i \sigma_2 \Delta_R \ell_{jR}+
(R \leftrightarrow L)\right)+h.c.
\label{treeY}
\end{eqnarray}
where $\tilde{\Phi} = \tau_2 \Phi^* \tau_2$. In the above Yukawa Lagrangian, the indices $i, j = 1, 2, 3$ correspond to the three families of fermions. The Majorana Yukawa couplings $f$ is same for both left and right handed neutrinos
because of left-right symmetry. These couplings $f$ give rise to the Majorana mass terms of both left handed and right handed neutrinos after the triplet Higgs fields $\Delta_{L,R}$ acquire non-zero vev. These mass terms appear in the seesaw formula as discussed below. The resulting seesaw formula in this minimal model can be written as
\begin{equation}
m_{LL}=m_{LL}^{II} + m_{LL}^I
\label{type2a}
\end{equation}
 where the usual  type I  seesaw formula  is given by the expression,
\begin{equation}
m_{LL}^I=-m_{LR}M_{RR}^{-1}m_{LR}^{T}
\end{equation}
Here  $m_{LR}$ is the Dirac neutrino mass matrix defined as $m_{LR} = y_{ij} \langle \Phi \rangle $. It should be noted that the Yukawa couplings $y_{ij}$ in the definition of Dirac neutrino mass matrix are not the same as the ones introduced in the Yukawa Lagrangian (\ref{treeY}), but the ones obtained at the electroweak scale after renormalization group evolution (RGE) effects are taken into account from the scale of left right symmetry breaking down to the electroweak scale. Such RGE effects on neutrino parameters for type I and type II seesaw models have been studied in \cite{rget1} and \cite{rget2} respectively. However, in our present work we do not attempt to perform a systematic RGE study of neutrino parameters in LRSM. To simplify our analysis, we assume that the RGE effects on the neutrino Yukawa couplings from the left right symmetry scale to the electroweak scale do not diminish the effects of higher dimensional Planck scale suppressed operators on neutrino mass in LRSM (as we discuss in next section).

In LRSM with Higgs triplets, $M_{RR}$ can be expressed as $M_{RR}=v_{R}f_{R}$ with $v_{R}$ being the vev of the right handed triplet Higgs field $\Delta_R$ imparting Majorana masses to the right-handed neutrinos and $f_{R}$ is the corresponding Yukawa coupling. The first term $m_{LL}^{II}$ in equation (\ref{type2a}) is due to the vev of $SU(2)_{L}$ Higgs triplet. Thus, $m_{LL}^{II}=f_{L}v_{L}$ and $M_{RR}=f_{R}v_{R}$, where $v_{L,R}$ denote the vev's and $f_{L,R}$ are symmetric $3\times3$ matrices. The left-right symmetry demands $f_{R}=f_{L}=f$. The induced vev for the left-handed triplet $v_{L}$ can be shown for generic LRSM to be
$$v_{L}=\gamma \frac{M^{2}_{W}}{v_{R}}$$
with $M_{W}\sim 80.4$ GeV being the weak boson mass such that 
$$ |v_{L}|<<M_{W}<<|v_{R}| $$ 
In general $\gamma$ is a function of various couplings in the scalar potential of generic LRSM and without any fine tuning $\gamma$ is expected to be of the order unity ($\gamma\sim 1$). The seesaw formula in equation (\ref{type2a}) can now be expressed as
\begin{equation}
m_{LL}=\gamma (M_{W}/v_{R})^{2}M_{RR}-m_{LR}M^{-1}_{RR}m^{T}_{LR}
\label{type2}
\end{equation}
\begin{figure}[h]
\begin{center}
\includegraphics[width=1.0\textwidth]{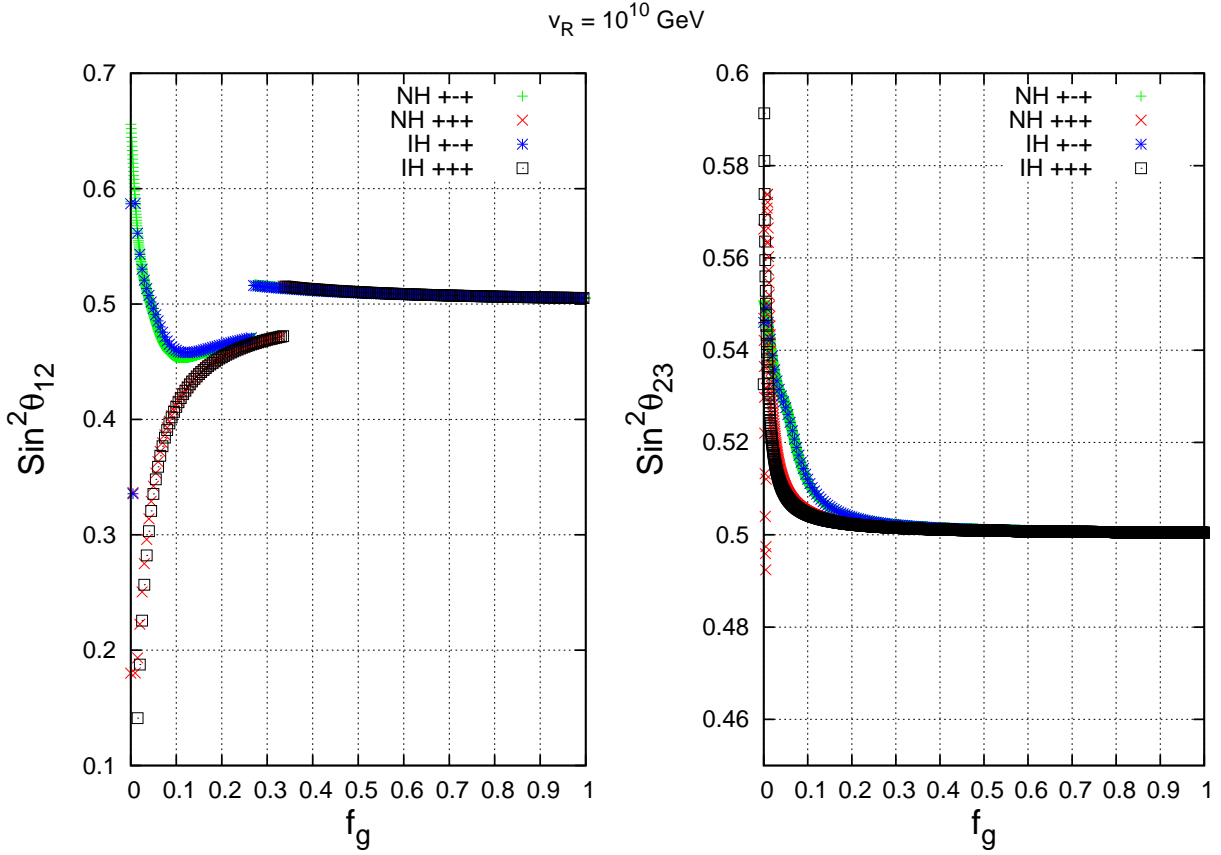}
\end{center}
\caption{Variations of $\text{sin}^2\theta_{12}$ and $\text{sin}^2\theta_{23}$ as a function of $f_g$ for $v_R = 10^{10}$ GeV}
\label{fig4}
\end{figure}
\begin{figure}[h]
\begin{center}
\includegraphics[width=1.0\textwidth]{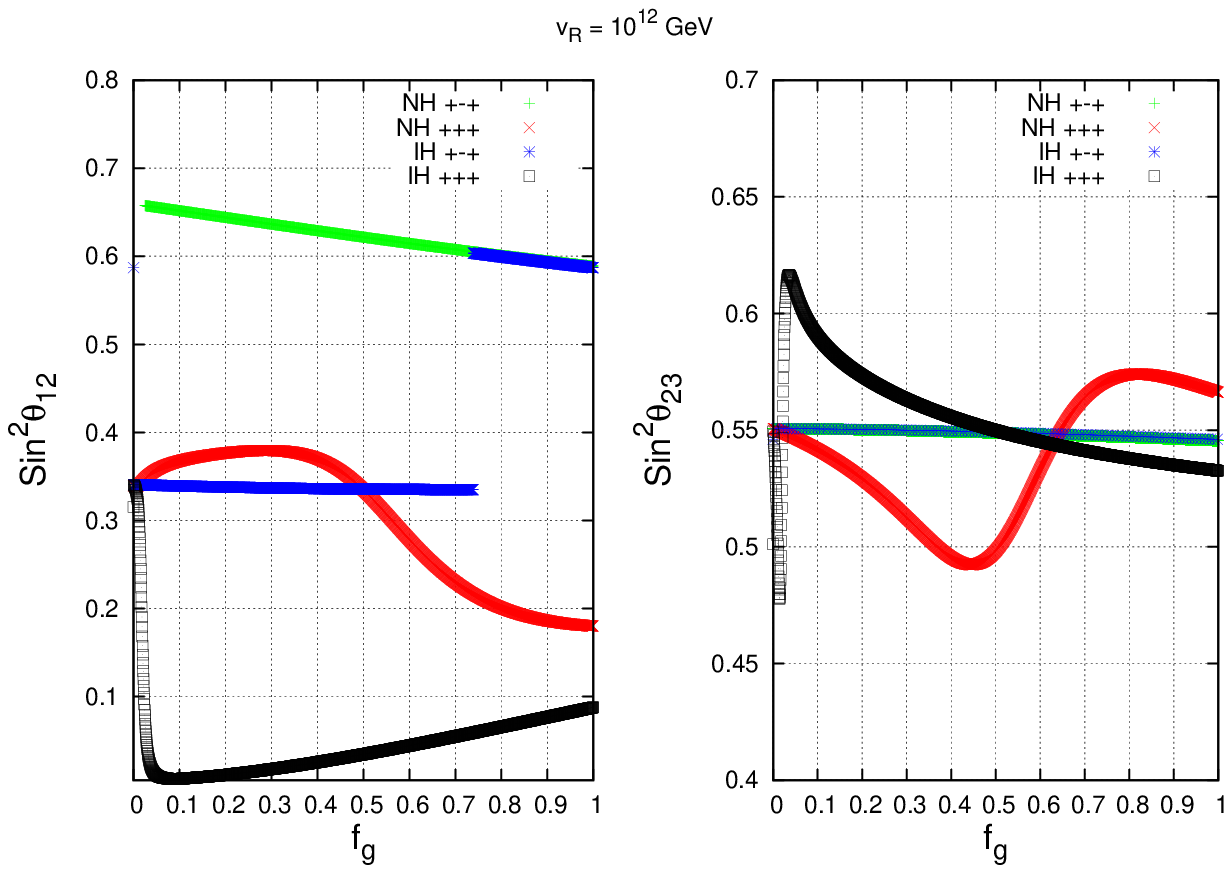} 
\end{center}
\caption{Variations of $\text{sin}^2\theta_{12}$ and $\text{sin}^2\theta_{23}$ as a function of $f_g$ for $v_R = 10^{12}$ GeV}
\label{fig5}
\end{figure}
\begin{figure}[h]
\begin{center}
\includegraphics[width=1.0\textwidth]{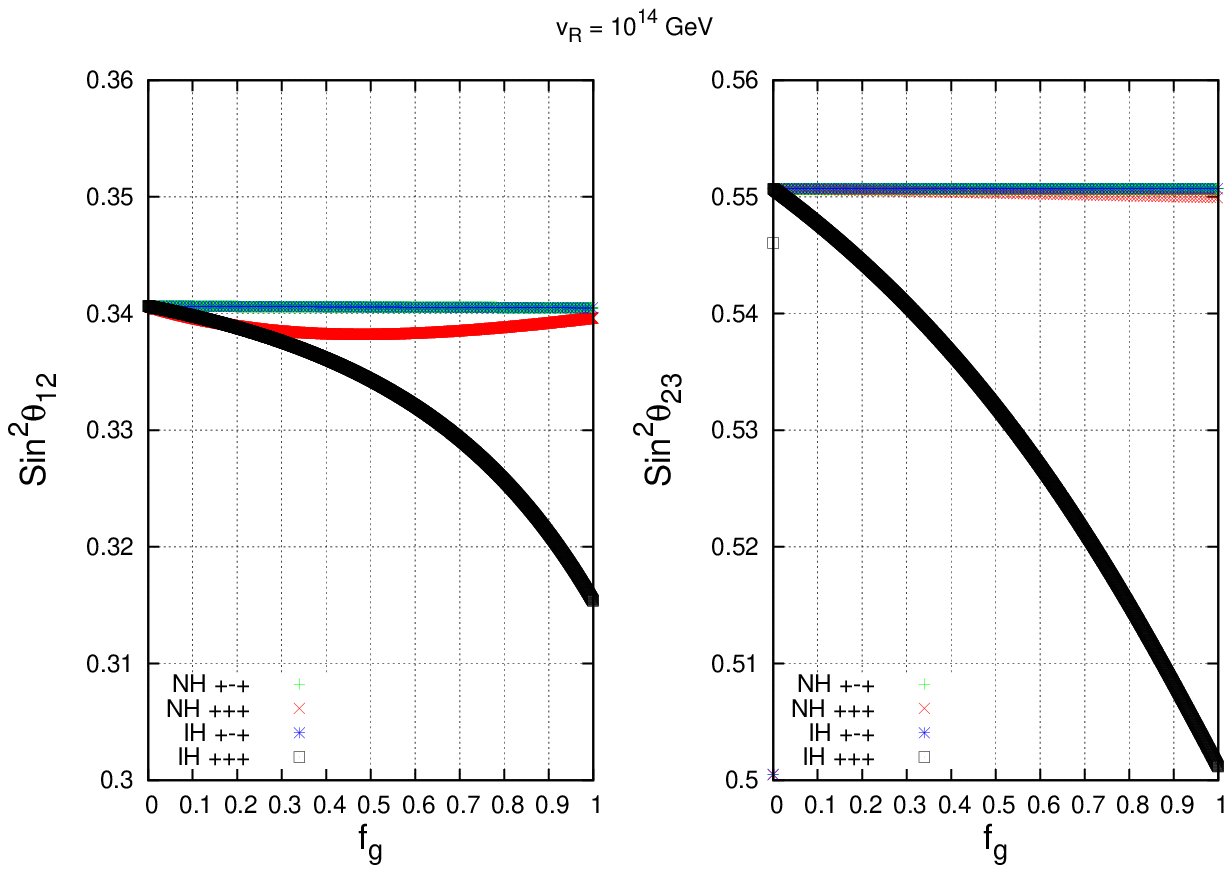}
\end{center}
\caption{Variations of $\text{sin}^2\theta_{12}$ and $\text{sin}^2\theta_{23}$ as a function of $f_g$ for $v_R = 10^{14}$ GeV}
\label{fig6}
\end{figure}

\section{Higher dimensional operators in LRSM}
\label{HDLRSM}
In the minimal LRSM discussed above, the next to leading order terms contributing to neutrino masses can be written as
\begin{equation}
{\cal L}^{NR} = f_{gL} \ell_{iR}^T \ C \ i \sigma_2 \Delta_R \ell_{jR} \frac{\Delta^{\dagger}_R \Delta_R}{M^2_{Pl}} + R \leftrightarrow L
\label{NLY}
\end{equation}
where $M_{Pl} \sim 10^{19}$ GeV is the Planck scale. Here $f_{gL} \ne f_{gR}$ and hence D-parity is explicitly broken with the introduction of the higher dimensional operators above. Now, using the tree level Yukawa terms (\ref{treeY}) as well as the Higher dimensional operators (\ref{NLY}), the right handed neutrino mass matrix can be written as
$$ M_{RR} = f v_R + f_{gR} \frac{v^3_R}{M^2_{Pl}} $$
The first term on the right hand side of equation (\ref{type2a}) takes the form
$$ m_{LL}^{II} = f v_L + f_{gL} \frac{v^3_L}{M^2_{Pl}} $$
$$ \Rightarrow m_{LL}^{II} = \gamma (M_{W}/v_{R})^{2} (f v_R + f_{gL} \frac{v^2_L v_R}{M^2_{Pl}}) $$
$$ \Rightarrow m_{LL}^{II} =  \gamma (M_{W}/v_{R})^{2} (M_{RR}+f_{gL} \frac{v^2_L v_R}{M^2_{Pl}} - f_{gR} \frac{v^3_R}{M^2_{Pl}} ) $$ 
Thus there arises two additional terms in the seesaw formula after the higher dimensional operators are taken into account. These two terms are proportional to $v^2_L/(v_RM^2_{Pl}) $ and $v_R/M^2_{Pl}$ respectively. We check that neither of these two correction terms can change the predictions of neutrino parameters from the ones predicted by the tree level seesaw formula (\ref{type2}). This is obviously because of the $1/M^2_{Pl}$ supression in both the terms which is almost negligible compared to the tree level neutrino mass terms.

Now, let us consider the presence of an additional gauge singlet field $\sigma$ in LRSM. Since a singlet like $\sigma (1,1,1,0)$ can naturally fit inside several $SO(10)$ representations, we assume the vev of this singlet field to be of order $\langle \sigma \rangle \sim M_{\text{GUT}} \sim  10^{16} \; \text{GeV} $. In the presence of such a field, the non-leading terms contributing to neutrino masses can be of dimension five as follows:
\begin{equation}
{\cal L}^{NR} = f_{gL} \ell_{iR}^T \ C \ i \sigma_2 \Delta_R \ell_{jR} \frac{\sigma}{M_{Pl}} + R \leftrightarrow L
\label{NLY2}
\end{equation}
Doing the same analysis as in the case of minimal LRSM, here $M_{RR}$ is found to be
$$ M_{RR} = f v_R + f_{gR} \frac{\langle \sigma \rangle v_R}{M_{Pl}} $$
The type II seesaw term $m_{LL}^{II}$ becomes 
$$ m_{LL}^{II} = \gamma (M_{W}/v_{R})^{2} (f v_R + f_{gL}\frac{\langle \sigma \rangle v_R}{M_{Pl}}) $$
$$ \Rightarrow m_{LL}^{II} = \gamma (M_{W}/v_{R})^{2} (M_{RR} +(f_{gL}-f_{gR}) \frac{\langle \sigma \rangle v_R}{M_{Pl}}) $$
Without losing any generality, we assume $(f_{gL}-f_{gR})$ to be a Hermitian matrix of order one multiplied a numerical factor $f_g$ which decides the overall strength of the corrected term. In the next section, we study the variation of neutrino mixing parameters as a function of this numerical factor $f_g$.

\section{Numerical Analysis}
\label{num}
The latest global fit value for $3\sigma$ range of neutrino oscillation parameters \cite{schwetz12} are as follows:
$$ \Delta m_{21}^2=(7.00-8.09) \times 10^{-5} \; \text{eV}^2$$
$$ \Delta m_{31}^2 \;(\text{NH}) =(2.27-2.69)\times 10^{-3} \; \text{eV}^2 $$
$$ \Delta m_{23}^2 \;(\text{IH}) =(2.24-2.65)\times 10^{-3} \; \text{eV}^2 $$
$$ \text{sin}^2\theta_{12}=0.27-0.34 $$
$$ \text{sin}^2\theta_{23}=0.34-0.67 $$ 
\begin{equation}
\text{sin}^2\theta_{13}=0.016-0.030
\end{equation}
where NH and IH refers to normal and inverted hierarchy respectively. Unlike the tight constraints on the above parameters, the global fit $3\sigma$ range for the value of Dirac CP phase $\delta_{CP}$ extends over entire $0-2\pi$ range. For illustrative purposes, here we take its value to be $300$ degrees (same as the central value given in \cite{schwetz12}).

\begin{figure}[h]
\begin{center}
\includegraphics[width=1.0\textwidth]{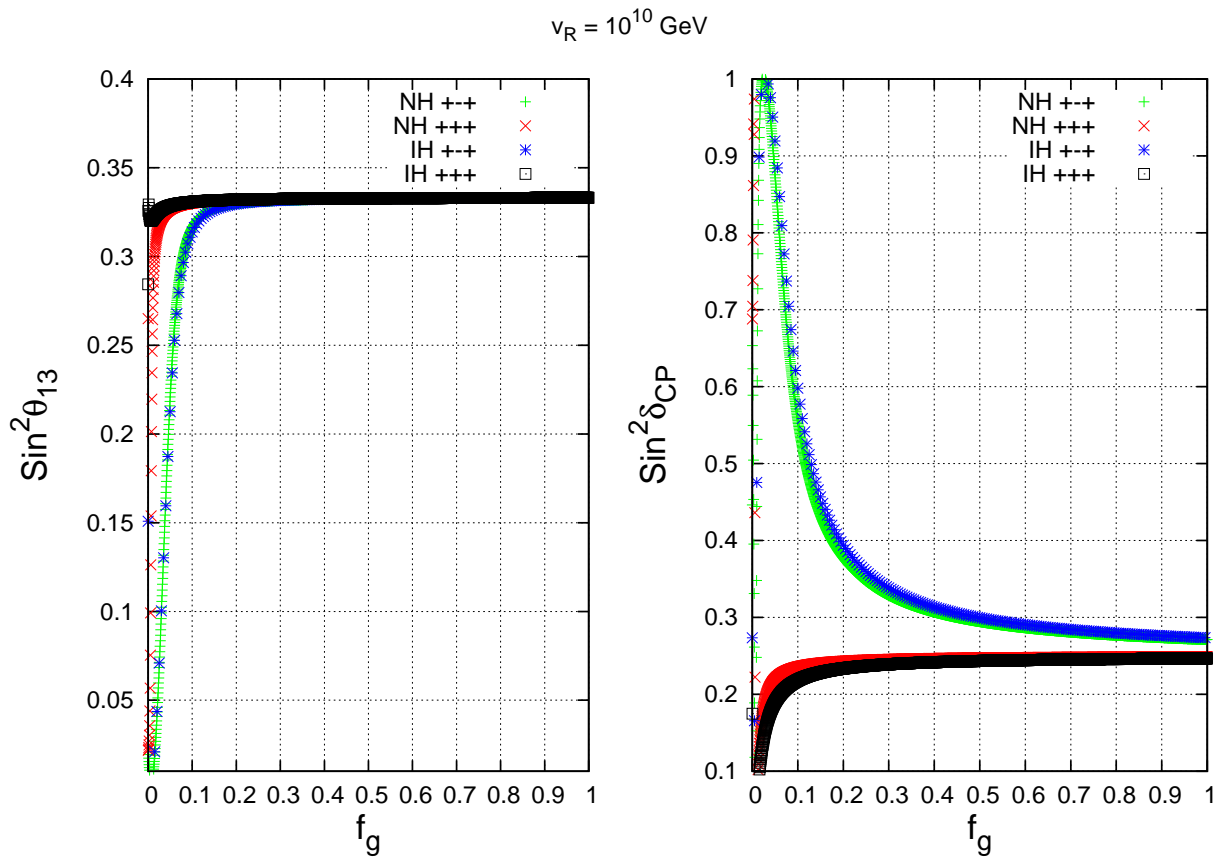} 
\end{center}
\caption{Variations of $\text{sin}^2\theta_{13}$ and $\text{sin}^2\delta_{CP}$ as a function of $f_g$ for $v_R = 10^{10}$ GeV}
\label{fig7}
\end{figure}
\begin{figure}[h]
\begin{center}
\includegraphics[width=1.0\textwidth]{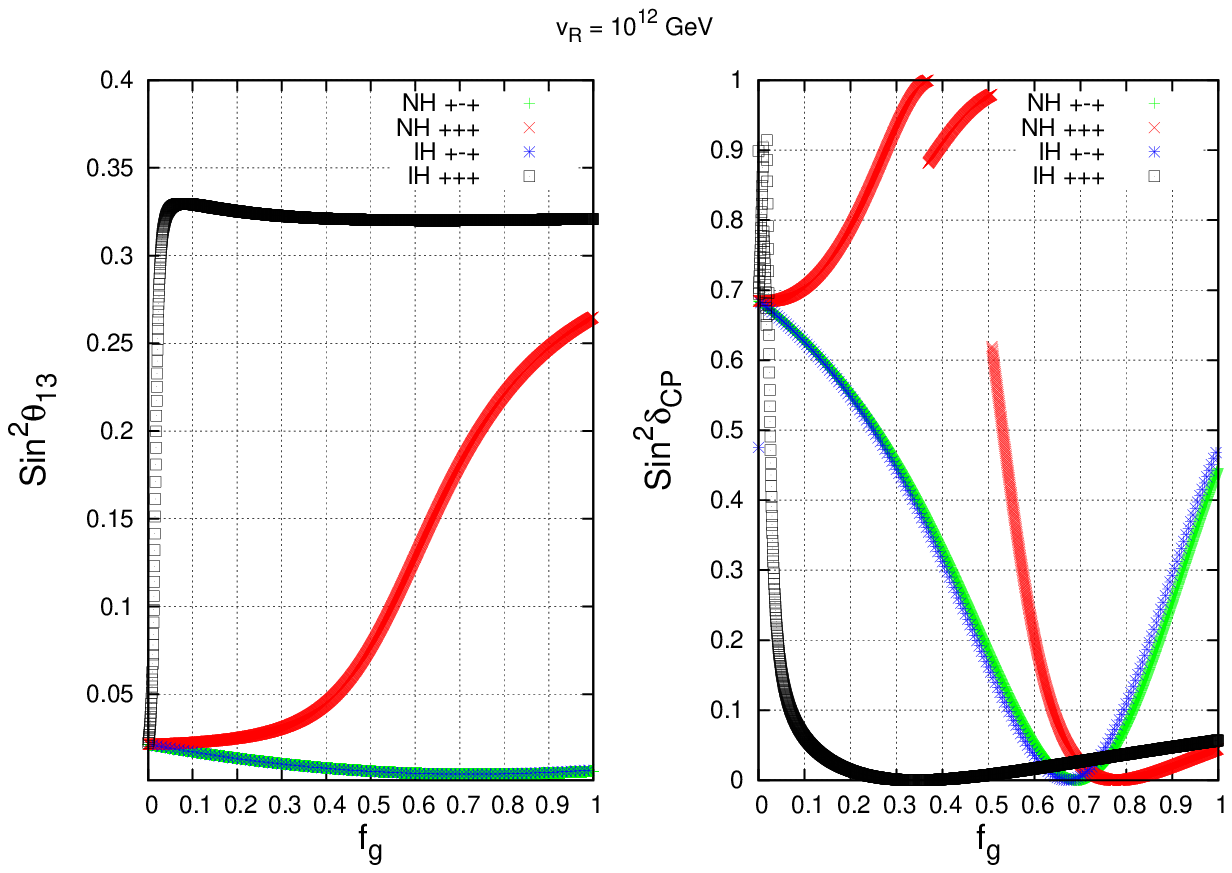} 
\end{center}
\caption{Variations of $\text{sin}^2\theta_{13}$ and $\text{sin}^2\delta_{CP}$ as a function of $f_g$ for $v_R = 10^{12}$ GeV}
\label{fig8}
\end{figure}
\begin{figure}[h]
\begin{center}
\includegraphics[width=1.0\textwidth]{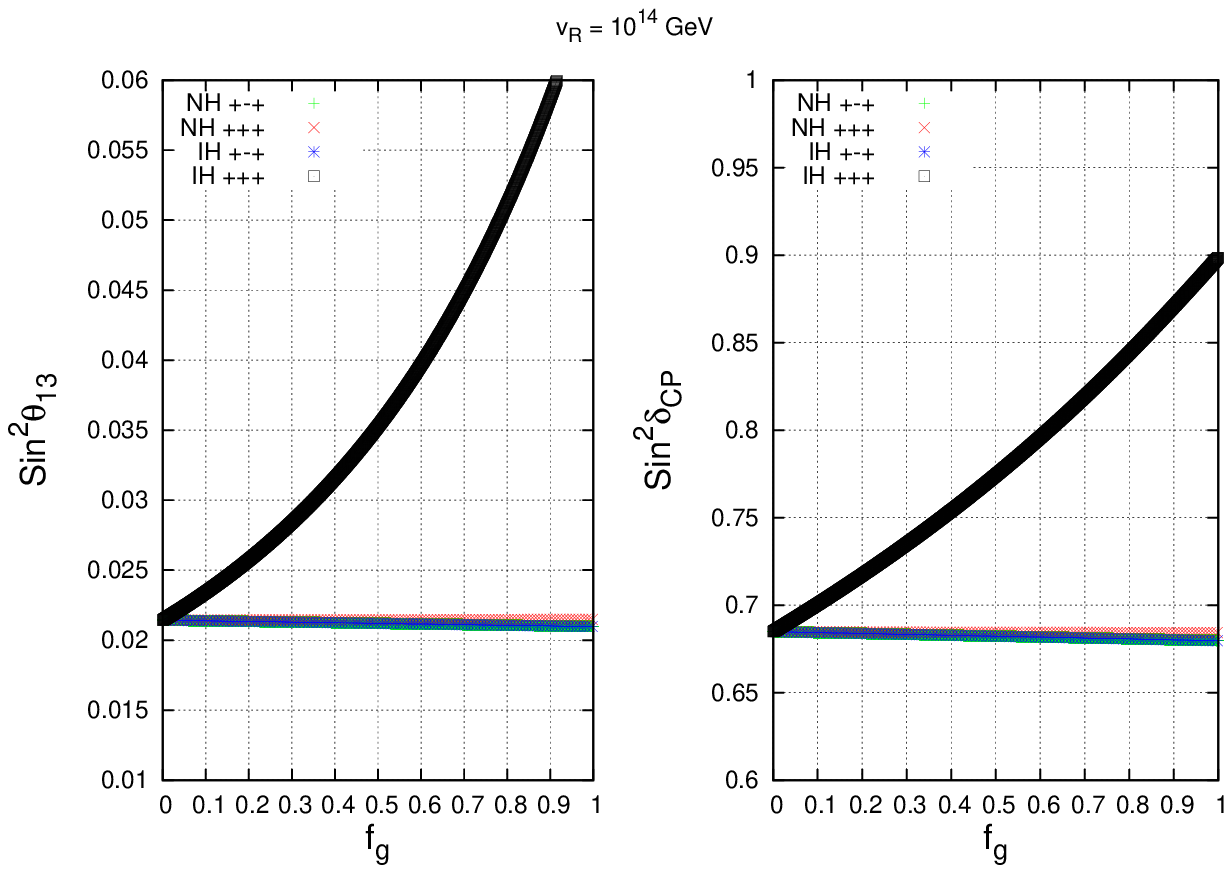}
\end{center}
\caption{Variations of $\text{sin}^2\theta_{13}$ and $\text{sin}^2\delta_{CP}$ as a function of $f_g$ for $v_R = 10^{14}$ GeV}
\label{fig9}
\end{figure}

For the purpose of our numerical analysis, we first fit the neutrino mass matrix $m_{LL}$ using the best fit global parameters mentioned above. For both normal and inverted hierarchical neutrino mass patterns, we consider extremal Majorana phases such that the mass eigenvalues are either $(m_1, m_2, m_3)$ or $(m_1, -m_2, m_3)$ denoted by $(+++)$ and $(+-+)$ respectively. We follow the same approach for numerical analysis as \cite{mkd-db-rm} where the variation of neutrino mixing parameters with respect to the dimensionless parameter $\gamma$ in the seesaw formula (\ref{type2}) was studied in details. 

After parametrizing the neutrino mass matrix for the tree level seesaw formula (\ref{type2}) using the global fit neutrino data, we introduce the correction term (\ref{NLY2}) to the seesaw formula. As discussed above, this correction term is of the form 
$$ m^{corr}_{LL} = \gamma (M_{W}/v_{R})^{2} (f_{gL}-f_{gR}) \frac{\langle \sigma \rangle v_R}{M_{Pl}} $$
Here we assume $(f_{gL}-f_{gR}) = f_g \mathcal{O}(1)$ where $\mathcal{O}(1)$ is a Hermitian matrix of order one. 

We then vary the dimensionless parameter $f_g$ from $10^{-5}$ to $1$ and see the variations of neutrino mixing parameters. The results are shown in figures \ref{fig1}, \ref{fig2}, \ref{fig3}, \ref{fig4}, \ref{fig5}, \ref{fig6}, \ref{fig7}, \ref{fig8} and \ref{fig9} for three different values of left-right symmetry breaking scales $v_R = 10^{10}, 10^{12}, 10^{14}$ GeV and both normal and inverted hierarchies as well as both types of extremal Majorana phases. As seen from the figures, the changes in the neutrino mixing parameters from the best fit values (corresponding to $f_g = 0$ in our case) become more and more significant as we go from $v_R = 10^{14}$ GeV to $v_R = 10^{10} $ GeV. In particular, for $v_R = 10^{14}$ GeV, almost all the neutrino parameters lie within the $3\sigma$ allowed range for all possible values of $f_g$. Only $\Delta m^2_{21}$ goes outside the $3\sigma$ range for $f_g > 0.1$ for all the models and $\theta_{13}$ deviates from the allowed range for $f_g > 0.35$ for IH(+++) model. For $v_R = 10^{12}$, the mass squared differences lie within the allowed range only when $f_g < 0.01$ whereas for $v_R = 10^{10}$ GeV, they lie outside the $3\sigma$ range for entire range of $f_g$ parameter under study. Similarly, the mixing angles are also found to lie within the allowed range only for some small range of parameter space for lower values of $v_R$.

\section{Results and Conclusion}
\label{conclude}
We have studied the effects of higher dimensional Planck scale suppressed operators on neutrino masses and mixings in left right symmetric extension of standard models. These higher dimensional correction terms arise due to the fact that any theory of quantum gravity does not respect global symmetries: both continouous and discrete. Since left right symmetric models have an in-built discrete global symmetry called D-parity, it is generic to introduce explicit D-parity breaking terms suppressed by the scale of gravity or the Planck scale. We have shown that in the minimal LRSM, the order of such higher dimensional operators have dimension at least six and hence too small to affect neutrino masses and mixing. We then incorporate the presence of additional gauge singlet scalar field which allows dimension five Planck suppressed operators contributing to the neutrino mass matrix. Such gauge singlet field can be naturally fit within several $SO(10)$ GUT multiplets. As discussed in our earlier work \cite{borah12}, these singlet scalar fields play a non-trivial role in destabilizing domain walls which arise in these models as a result of spontaneous D-parity breaking. 

Sticking to the issue of neutrino mass alone in the present work, we then fit the tree level neutrino mass matrix to the global best fit neutrino data. After doing this, we introduce the higher dimensional operators and see the variations in the neutrino mixing parameters with the changes in the overall coupling strength of these opearators. We consider both normal as well as inverted hierarchies and two extremal Majorana phases in our work. Doing this exercise for three different left right symmetry breaking scales namely $10^{14}, 10^{12}, 10^{10}$ GeV, we show that the effects of these operators can be very significant for those models with left right symmetry breaking scale below $10^{14}$ GeV. It should be noted that the purpose of our study is not to rule out or disfavor any particular model, but to emphasize the fact that fitting the tree level seesaw formula with neutrino data is not enough in these models. The higher dimensional opearators which violate D-parity explicitly, can give rise to sizable contributions and hence have to be taken into account in generic left right symmetric models.

\end{document}